\documentclass[aps,prl,twocolumn,groupedaddress,showpacs]{revtex4-1}
\usepackage{graphicx}
\usepackage{amsmath}
\usepackage{textcomp}
\usepackage{mathrsfs}
\usepackage{amsfonts}

\begin{document}

\newcommand{\beq}{\begin{equation}}
\newcommand{\eeq}{\end{equation}}
\newcommand{\barr}{\begin{eqnarray}}
\newcommand{\earr}{\end{eqnarray}}
\newcommand{\bseq}{\begin{subequations}}
\newcommand{\eseq}{\end{subequations}}
\newcommand{\vett}[1]{\mathbf{#1}}
\newcommand{\uvett}[1]{\hat{\vett{#1}}}
\newcommand{\mat}[4]{\left[
\begin{array}{cc}
#1 & #2 \\ #3 & #4 \\ 
\end{array}
\right]}

\title{The Effect of Orbital Angular Momentum on Nondiffracting Ultrashort Optical Pulses}

\author{Marco Ornigotti$^1$}

\email{marco.ornigotti@uni-jena.de}

\author{Claudio Conti$^{2,3}$}
\author{Alexander Szameit$^1$}

\affiliation{$^1$Institute of Applied Physics, Friedrich-Schiller Universit\"at Jena, Max-Wien Platz 1, 07743 Jena, Germany}
\affiliation{$^2$ Institute for Complex Systems (ISC-CNR), Via dei Taurini 19, 00185, Rome, Italy}
\affiliation{$^3$  Department of Physics, University Sapienza, Piazzale Aldo Moro 5, 00185, Rome, Italy}

\date{\today}

\begin{abstract}
We introduce a new class of nondiffracting optical pulses possessing orbital angular momentum. 
By generalizing the X-waves solution of the Maxwell equation, we discover the coupling between angular momentum and the temporal degrees of freedom of ultra-short pulses.
The spatial twist of propagation invariant light pulse turns out to be directly related to the number of optical cycles.
Our results may trigger the development on novel multi-level classical and quantum transmission channels free of dispersion and diffraction,  
may also find application in the manipulation of nano-structured objects by ultra-short pulses, and for novel approaches to the spatio-temporal measurements in ultrafast photonics.
 \end{abstract}


\maketitle
Since the development of the laser, and in particular of mode locking \cite{modeL} and Q-switching \cite{Qsw} techniques, optical pulses have attracted a lot of interest, and their development influenced many fields of fundamental and applied research such as atomic physics, spectroscopy, communications, material processing and medicine, to name a few \cite{OP1,OP2}. As they are essentially suitable superpositions of plane waves that travel with different frequencies, optical pulses tend to diffract in both space and time during propagation. This feature constitutes a limiting factor for some applications like lithography \cite{lito}, where the spatial broadening affects the quality of the generated mask, or communication science, where temporal broadening can induce additional noise between adjacent channels \cite{agrawal}. To solve these issues, so-called localized waves \cite{localizedWaves}, i.e., nondiffracting electromagnetic fields in both space and time, have been proposed in the last decades, with their most famous representatives being the X-waves. First introduced in acustics \cite{ref3,ref4},  X-waves were subsequently studied in different areas of physics, such as nonlinear optics \cite{ref5,ref6}, Bose-Einstein condensates \cite{ref7}, quantum optics \cite{ref8} and waveguide arrays \cite{ref8bis1,ref8bis}, to name a few. Recently, they have also been proposed as a possible solution to efficient free space communication \cite{freeSpace}. 

X-waves are polychromatic superpositions of Bessel waves, and the related vast literature  mostly considers superpositions of Bessel beams of order zero, and neglects their possible orbital angular momentum (OAM) content \cite{OAMX}. OAM is indeed related to the twisted wavefront of higher order Bessel beam solutions of the Maxwell equations \cite{libroOAM}. 
Seemingly, despite some published papers \cite{pulseAM1, pulseAM2}, OAM is still seldomly  associated to ultrashort pulses, and only very recently some experimental results reported
femtosecond vortex beams \cite{pulseOAM1,pulseOAM2} and Laguerre-Gauss supercontinuum \cite{pulseOAM3}. 

The fact that light possesses both linear and angular momentum is known since the pioneering works by Poynting \cite{poynting} and Darwin \cite{darwin}. However, it is only thanks to the seminal works of Berry and Nye in 1974 \cite{berry} and Allen and Woerdman in 1992  \cite{allenWoerdman} that the OAM concept was brought into the field of optical beams, where it has been extensively studied both from a fundamental \cite{sep1,sep2,sep3,sep4} and experimental point of view, leading to striking applications such as optical tweezers \cite{tweezer} and spanners \cite{spanner}. Very recently, OAM has also been proposed as a new degree of freedom for encoding information in a superdense manner, and both its classical \cite{refCoding1} and quantum \cite{refCoding2} features have been investigated. 

A challenging issue at the moment is generating ultrashort pulses with variable OAM; this would open unprecedented possibilities in terms of lightwave transmission systems unaffected by dispersion
and diffractions. In this terms, a very prosiming direction is combining the non-diffracting character of X-waves with the superdense coding by OAM. Moreover, the extension of the concept of OAM to the domain of ultra-short optical pulses 
may give new insights on light-matter interaction, as recent works suggest \cite{vortexIon}.

Following the recent experimental realization of higher order Bessel beams by holographic techniques \cite{BesselHigh}, in this Letter we propose a new class of OAM-carrying non diffracting pulses. 
We consider superpositions of $m$-th order Bessel beams with an exponentially decaying spectrum, and generalize the well known fundamental X-waves \cite{localizedWaves}. 
This new class of few-cycles optical pulses is not only an exact model for the connection between OAM and the ultra-fast photonics, but they are a new tool for OAM-based free-space quantum and classical communications by localized waves.

\begin{figure}[!t]
\begin{center}
\includegraphics[width=0.5\textwidth]{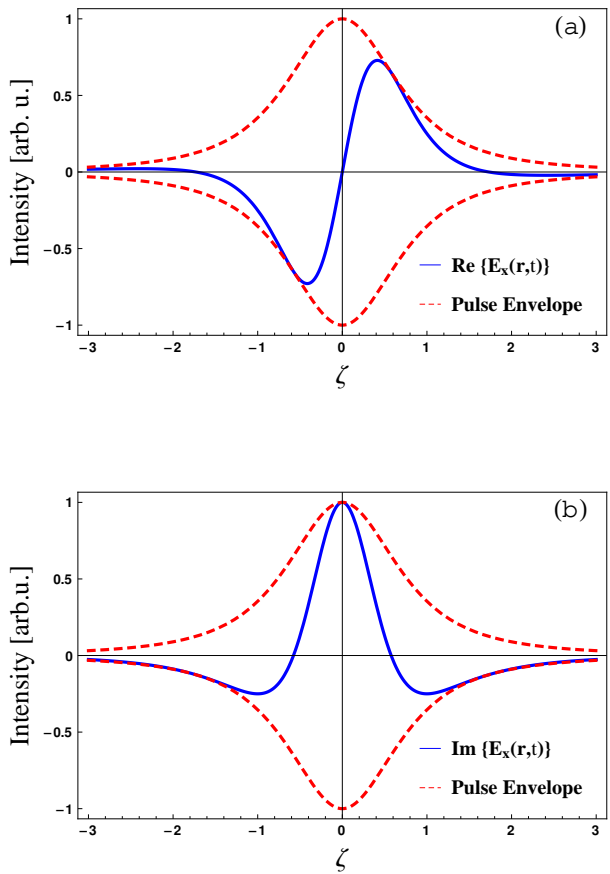}
\caption{Real (a) and imaginary (b) part of  the electric field component $E_x(\vett{r},t)$ as a function of the co-moving coordinate $\zeta=(z/c)\cos\vartheta_0-t$, for the case $m=1$. The real part is an odd function that crosses the line $\zeta=0$ three times, while the imaginary part is an even function, with two crossings. Parameters: $\vartheta_0=0.01$ (corresponding to the paraxial case) and $\alpha=1$; $\zeta$ is dimensionless.}
\label{figure1}
\end{center}
\end{figure}
%
We first introduce the general solution for X-waves with OAM.  We start from a Bessel beam \cite{durnin}, namely:
\beq\label{eq1}
\psi(R,\theta,z)=\text{J}_m\left(\frac{\omega}{c}\sin\vartheta_0 R\right)e^{im\theta}e^{i\omega z\cos\vartheta_0},
\eeq
where $\omega=ck$ and $\vartheta_0$ is the Bessel cone angle, i.e., the beam's characteristic parameter. Bessel beams carry OAM because they are eigenstates of the $z$-components of the angular momentum operator $\hat{L}_z=-i\partial/\partial\theta$ with eigenvalue $m$ \cite{AMQM}. It is well known that Eq. \eqref{eq1} well describes a monochromatic field. It is not difficult, however, to generalize this result to the non-monochromatic case, where, following Ref. \cite{localizedWaves}, the scalar field
\beq\label{phi}
\phi(\vett{r},t)=e^{im\theta}\int d\omega\; f(\omega)\text{J}_m\left(\frac{\omega}{c}\sin\vartheta_0 R\right)e^{-i\omega (t-\frac{z}{c}\cos\vartheta_0)},
\eeq
is an exact solution of the full wave equation, being $f(\omega)$ an arbitrary spectrum. Equation \eqref{phi} is known in literature to describe localized waves, namely field configurations that are well localized both in space and time. If the following exponentially decaying spectrum is used
\beq\label{spectrum}
f(\omega)=\omega^ne^{-\alpha\omega}\Theta(\omega),
\eeq
where $\alpha$ is a parameter with the dimensions of a time that controls the width of the spectrum and $\Theta(\omega)$ is the Heaviside step function \cite{nist}, Eq. \eqref{phi} with $m=0$ describes the well known \emph{fundamental X-waves} \cite{localizedWaves}.  For $m\neq 0$, however, Eq. \eqref{phi} admits the following analytical solution \cite{gradstein}:
\barr\label{integralJ}
\phi_m(\vett{r},t)=\frac{C_{n,m}\xi^me^{im\theta}}{(\alpha+i\zeta)^{n+1}} \, _2F_1\left(a,b;m+1;-\xi^2\right),
\earr
where $C_{n,m}=(n+m)!/(2^mm!)$, $a=(n+m+1)/2$, $b=(n+m+2)/2$, $\xi=R\sin\vartheta_0/[c(\alpha+i\zeta)]$, $\zeta=(z/c)\cos\vartheta_0-t$ is the co-moving reference frame attached to the X-wave and $\, _2F_1(a,b;c,z)$ is the Gauss hypergeometric function \cite{nist}.  Eq. \eqref{integralJ} represents a scalar X wave carrying $m$ units of OAM. 
Moreover, as ultrashort pulses are often modeled through an exponentially decaying spectrum \cite{OP1}, Eq. \eqref{integralJ} can be therefore taken as a scalar ultrashort non diffracting wave that carries OAM. To correctly describe an optical ultrashort pulse, however, the scalar theory is no longer sufficient, and a full vector theory is required. An exact vectorial solution of Maxwell's equation can be built from a scalar function by means of the so-called Hertz vector potentials \cite{stratton}. 
We choose $\boldsymbol\Pi(\vett{r},t)=\uvett{z}\;\phi_m(\vett{r},t)$ as the Hertz potential, and the electric and magnetic vector fields are given by \cite{stratton}:
\bseq\label{hertzEq}
\begin{align}
\vett{E}(\vett{r},t)&=\nabla\times\nabla\times\boldsymbol\Pi(\vett{r},t),\\
\vett{B}(\vett{r},t)&=\frac{\partial}{\partial t}\left[\nabla\times\boldsymbol\Pi(\vett{r},t)\right].
\end{align}
\eseq

\begin{figure}[!t]
\begin{center}
\includegraphics[width=0.5\textwidth]{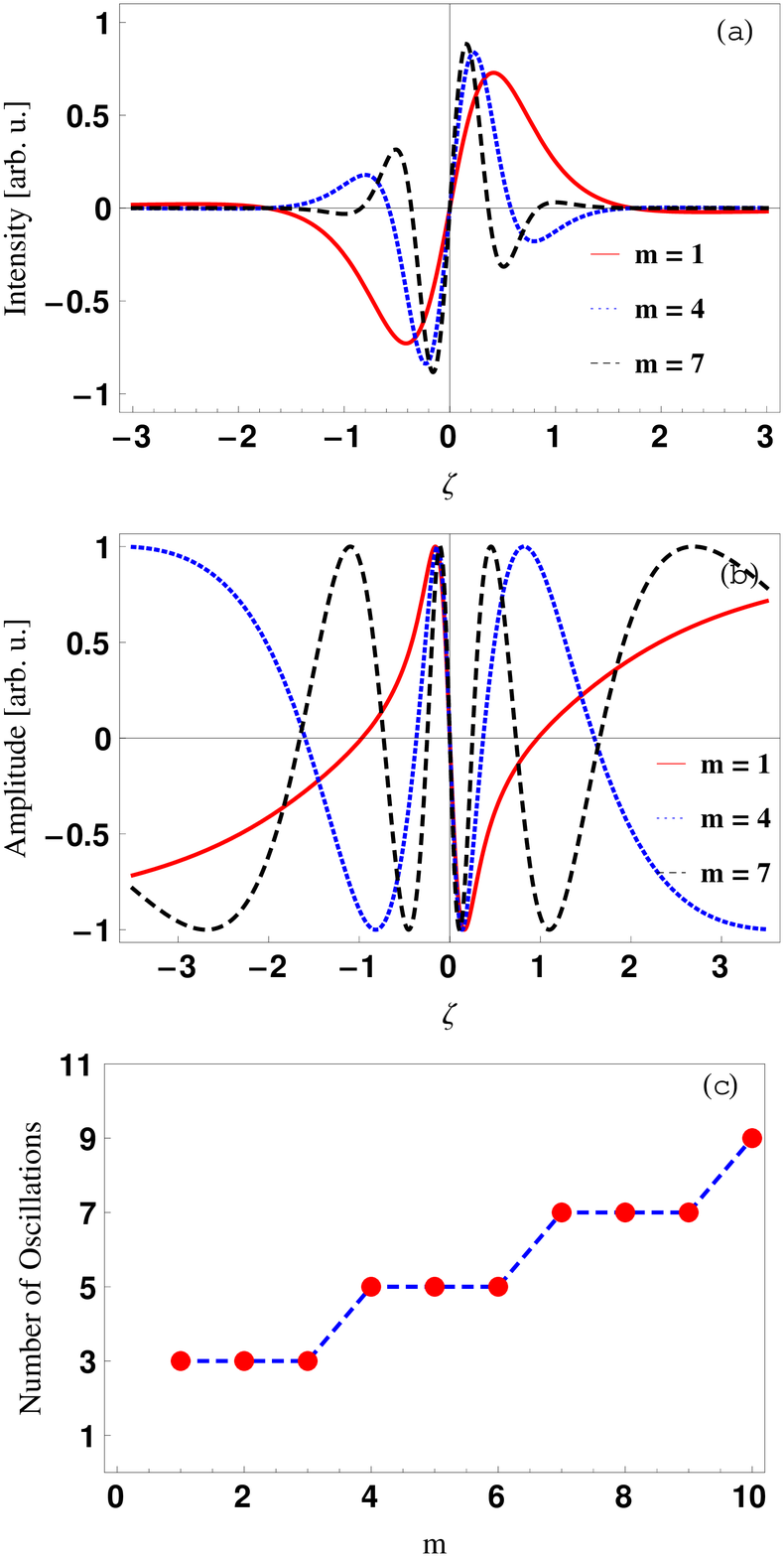}
\caption{(a) Real part of the $x$-component of the electric field $E_x(\vett{r},t)$ for different values of the OAM parameter $m$. (b) Oscillatory function $\cos\psi(\vett{r},\zeta)$ for different values of the OAM parameter. The choice of these values corresponds to $\omega_c=3/\alpha$, $\omega_c=6/\alpha=2(\omega_c)|_{m=1}$ and $\omega_c=9/\alpha=3(\omega_c)|_{m=1}$, respectively. When the OAM increases, the carrier frequency blue shifts, with an increase of the number of optical cycles, but the envelope FWHM decreases, thus resulting in a shorter pulse. (c) Calculated (red dots) number of oscillations performed by $\cos\psi(\vett{r},\zeta)$ as a function of $m$. 
Every three units of $m$ added, the number of cycles of $\cos\psi(\vett{r},\zeta)$ grows by one unity. The number of oscillations is calculated by counting the number of zero crossing $N_0$ of each cosine term and using $N_0=2N_{osc}+1$. Here, $\vartheta_0=0.01$ (corresponding to the paraxial case) and $\alpha=1$ has been chosen. In particular, $\alpha=1$ means that $\zeta$ is a dimensionless quantity.}
\label{figure2}
\end{center}
\end{figure}

The explicit expression of the vector electric and magnetic fields generated with the above equations are available in the Supplementary Material. The real and imaginary parts of the $x$-component of the electric field $E_x(\vett{r},t)$ are depicted in Fig. \ref{figure1} as a function of the co-moving coordinate $\zeta$. As can be seen, the real part is an odd function that crosses the axis $\zeta=0$ three times, while the imaginary part of the field is an even function, with only two crossings. 
This corresponds to a single-cycle optical-pulses following the definition in Ref. \cite{singleCycle}.  Note, moreover, that single-cycle optical-pulses are often described using of the so-called Ziolkowski's \emph{modified power spectrum} solution \cite{ziolkowski} as generating scalar function. Here, instead, we have used fundamental X-waves with OAM to generate a new class of diffractionless single cycled optical pulses that naturally carry OAM. This is the first result of this Letter: Eq. \eqref{integralJ} and its vector counterpart, defined through Eqs. \eqref{hertzEq}, represent a new class of optical pulses, namely fundamental X-waves with OAM. With this result we can investigate the effects of OAM on optical pulses. 

We consider, in particular, the case $m=1$, i.e., we assume to have a single-cycled optical pulse with one unit of OAM, and we study the connection between OAM and its temporal properties. We focus our attention on the $E_x(\vett{r},t)$ component; a similar discussion also holds for the other components. Fig. \ref{figure2}(a) shows the real part of the field for various values of the OAM parameter $m$. As can be seen, OAM affects the pulse's temporal properties  by changing its carrier frequency.
\begin{figure}[!t]
\begin{center}
\includegraphics[width=0.5\textwidth]{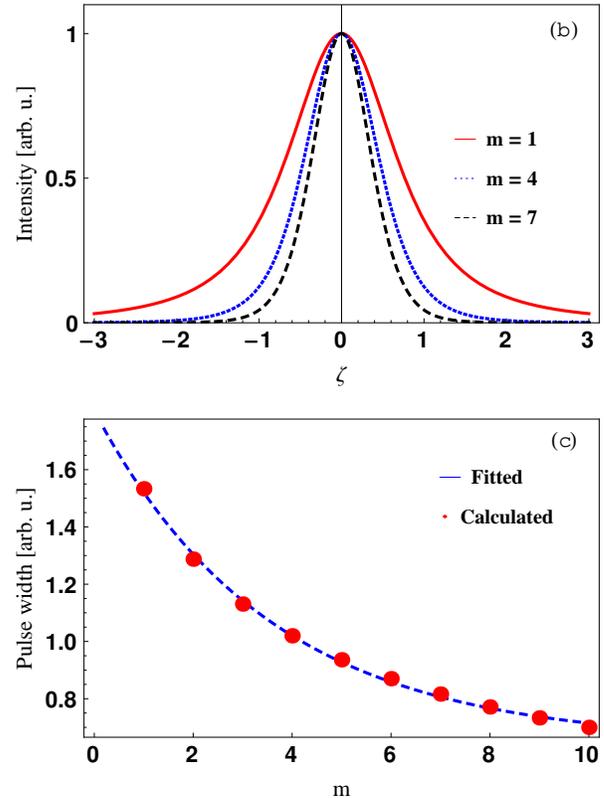}
\caption{(a) Pulse envelope of the $x$-component of the electric field $E_x(\vett{r},t)$ for different values of the OAM parameter $m$. (b) Calculated (red dots) and fitted (blue dashed line) pulse width $\Delta\tau$ as a function of the OAM $m$. The fit predicts an exponential shortening of the pulse width with respect to $m$, according to $Ae^{-Bm}+C$. We have $A=1.16$, $B=0.28$ and $C=0.64$. Here, $\vartheta_0=0.01$ (corresponding to the paraxial case) and $\alpha=1$ has been chosen. In particular, $\alpha=1$ means that $\zeta$ is a dimensionless quantity.  }
\label{figure3}
\end{center}
\end{figure}
As $m$ grows, in fact, the pulse's carrier frequency $\omega_c$ also increases. Correspondigly, the field oscillates more rapidly, and the number of the optical cycles changes. To estabilish the relation between the carrier frequency $\omega_c$ 
and the OAM parameter $m$, we recall that an optical pulse is written as the product of an envelope function $A(\vett{r},t)$ and an harmonic term, i.e., $A(\vett{r},t)\exp{(-i\omega_c t)}$. As detailed in the supplementary material, for a general field, the carrier frequency can be calculated as the derivative of the phase $\psi$ of the field in $t=0$. 
Here, $\psi(\vett{r},t)=\arg[E_x(\vett{r},t)]$ and we obtain:
\beq\label{omegaC}
\omega_c=\left|\frac{\partial\psi(\vett{r},\zeta)}{\partial\zeta}\right |_{\zeta=0}=\frac{m+2}{\alpha},
\eeq
where $\alpha$ is the spectral width of the pulse and the derivative has been taken with respect to the co-moving coordinate $\zeta$. It is worth noticing that the result of Eq. \eqref{omegaC} is exact only in the paraxial regime, where $\vartheta_0\ll 1$. However, although for the nonparaxial case the expression of $\omega_c$ is much more complicated, it can be demonstrated that the variation of $\omega_c$ with $m$ can still be well described by a slightly modified version of Eq. \eqref{omegaC}, namely $\omega_c=(m+2)/\alpha+\sigma_m$, where $\sigma_m$ accounts for the nonparaxial corrections. Although $\sigma_m$ actually depends on $m$, this dependence is very weak, and it can be treated, at the leading order in $m$, like a constant shift. Eq. \eqref{omegaC} is therefore valid independently of the value of $\vartheta_0$. 

Fig. \ref{figure2}(b) shows the oscillatory term  $\cos\psi(\vett{r},\zeta)$ (which, apart from an unimportant multiplicative factor, represents the real part of the field $E_x(\vett{r},t)$ \cite{note}) for various values of the OAM parameter $m$). This term increases by one optical cycle every three units of OAM. This result can be interpreted in a simple intuitive way: as the amount of OAM carried by the pulse grows, the pulse itself adapts to it by increasing the number of optical cycles that it is able to perform under its envelope.

The additional effect of OAM on the pulse is reported in Fig. \ref{figure3}(a), where the envelope of the field is plotted for various values of the OAM parameter $m$ . As the amount of OAM increases, the pulse duration given by its full-width half-maximum (FWHM) $\Delta\tau$, becomes smaller. To quantify this OAM-driven compression, we have numerically estimated $\Delta\tau$ for various $m$, and we show the results in Fig. \ref{figure3} (b): 
$\Delta\tau$ decreases exponentially with $m$. 
This is the second result of our Letter. The amount of OAM carried by a nondiffracting optical pulse affects its temporal properties, namely it varies its time duration $\Delta\tau$ (making it smaller as $m$ increases) and also changes its carrier frequency in such a way that the pulse gains an extra optical cycle every three units of OAM it carries (after the first one). In other words, in order to have high values of OAM in propagation-invariant pulses one needs to resort to higher frequencies and shorter temporal duration. At a fixed carrier frequency the maximal angular momentum is given by Eq. \eqref{omegaC}.

In conclusion we have introduced a new class of optical pulses possessing OAM by generalizing the well-known fundamental X-waves. We have theoretically investigated the effects of OAM onto such pulses and shown that, as the amount of OAM carried by the pulse increases, the pulse's carrier frequency increases accordingly, thus resulting in a shortening of the pulse width and the appearance (at certain discrete threshold of OAM) of extra field oscillations.  Given the enormous interest that OAM has generated  in the past years, we believe that its extension to the domain of optical pulses presented here with OAM-carrying X-waves will open the way for new and intriguing applications. In particular, this result sets a fundamental limit to the amount of OAM that a non diffracting optical pulse can carry. This could have interesting consequences for the case of superdense free space communication protocols with X-waves.

The authors thank the German Ministry of Education and Science (ZIK 03Z1HN31) for financial support.


\begin{thebibliography}{99}
\bibitem{note} In the complex exponential notation, the electric field can be written as $E_x(\vett{r},t)=\rho(\vett{r},t)\exp{[i\psi(\vett{r},t)]}$, where $\rho(\vett{r},t)=\left|E_x(\vett{r},t)\right|$. Then, the real part of this field can be written as 
$Re\{E_x(\vett{r},t)\}=\rho(\vett{r},t)\cos\psi(\vett{r},t)$. To analyze the frequency properties of this field, it is therefore enough to analyze only its oscillatory part, i.e., $\cos\psi(\vett{r},t)$.
\end{thebibliography}
\end{document}